\documentclass[12pt]{article}

\usepackage{amssymb}

\usepackage{verbatim}

\def\bS {{\mathbb{S}}}
\def\bR {{\mathbb{R}}}
\def\bN {{\mathbb{N}}}
\def\bC {{\mathbb{C}}}
\def\bH {{\mathbb{H}}}

\def\bZ {{\mathbb{Z}}}

\def\cC {{\mathcal C}} 

\def\cN {{\mathcal N}} 

\def\cS {{\mathcal S}}

\def\Di {\displaystyle}

\def\MR {\mathrm }

\def\sp {\mathrm {sp}}
\def\mb {{\bf b}} 
\def\mbt {{\widetilde {\bf b}}}
\def\mM {{\bf M}}

\makeatletter
\@addtoreset{equation}{section}

\makeatother

\newtheorem{theorem}{Theorem}[section]
\newtheorem{lemma}[theorem]{Lemma}
\newtheorem{proposition}[theorem]{Proposition}

\newtheorem{remark}[theorem]{Remark}
\newtheorem{corollary}[theorem]{Corollary}

\begin{document}

\bibliographystyle{plain}

\begin{center}
{\Large \bf {Magnetic bottles on geometrically finite hyperbolic surfaces}}
\end{center}

\vskip 0.5cm

\centerline{\bf { Abderemane MORAME$^{1}$
 and  Fran{\c c}oise TRUC$^{2}$}}

{\it {$^{1}$ Universit\'e de Nantes,
Facult\'e des Sciences,  Dpt. Math\'ematiques, \\
UMR 6629 du CNRS, B.P. 99208, 44322 Nantes Cedex 3, (FRANCE), \\
E.Mail: morame@math.univ-nantes.fr}}

{\it {$^{2}$ Universit\'e de Grenoble I, Institut Fourier,\\
            UMR 5582 CNRS-UJF,
            B.P. 74,\\
 38402 St Martin d'H\`eres Cedex, (France), \\
E.Mail: Francoise.Truc@ujf-grenoble.fr }}

\begin{abstract}
We consider a  magnetic Laplacian  
 $-\Delta_A=(id+A)^\star (id+A)$\\  
 on a hyperbolic surface  $\mM , $ 
 when the magnetic field $dA$ is infinite at the boundary at infinity. 
  We prove that   the counting function 
  of the eigenvalues has a particular  asymptotic behavior when 
$\mM$    has an infinite area. 
\footnote{ {\sl Keywords}~: spectral asymptotics,  
  magnetic bottles, hyperbolic surface.}
\end{abstract}

\section{Introduction}

We consider  a smooth, connected, complete and oriented Riemannian surface 
$(\mM, g)$ and  a smooth, real 
 one-form $A$ on $\mM  .$  We  define the magnetic Laplacian 
\begin{equation}\label{DeA} 
\begin{array}{c} 
- \Delta_A\; =\; (i\ d + A)^\star (i\ d +A)\; ,\\ 
 \left ( \; 
 (i\ d+A)u=i\ du+uA\; , \ \forall \; u\; \in \; C^\infty_0(\mM ; \bC )\; \right )\; . 
 \end{array} 
 \end{equation} 
The magnetic field is the exact two-form 
$\ \rho_B\; =\; dA\; .$\\ 

If $dm\; $ is the Riemannian measure on $\mM \; ,$ then 
\begin{equation}\label{DeMb} 
\rho_B\; =\; \mbt \; dm\; , \quad \MR{with} \quad \mbt \; \in \; 
C^\infty (\mM ;\bR )\; . 
\end{equation} 
The magnetic intensity is $\ \mb\; =\; |\mbt |\; .$ 

It is well known, (see \cite{Shu}  ), that $-\Delta_A$ has a unique 
self-adjoint extension on $L^2(\mM )\; ,$ containing in its domain 
$C_0^\infty (\mM ;\bC )\; ,$ the space of 
smooth and compactly supported functions. 

When $\; \mb\; $ is infinite at the infinity, (with some additional assumption), 
 the spectrum of $-\Delta_A\; $ 
is discrete, and we denote by $(\lambda_j)_j$ the increasing sequence of eigenvalues 
of $-\Delta_A\; ,$ (each eigenvalue is repeated according to its multiplicity). 
Let 
\begin{equation}\label{DeNL} 
N(\lambda )\; =\; \sum_{\lambda_j < \lambda } 1\; .
\end{equation} 

We are interested by the hyperbolic surfaces $\mM  ,$ when the curvature of 
$\mM  $   is  constant and negative. 

In this case, when $\mM  $   has finite area, the asymptotic behavior of $\; N(\lambda )$ 
seems to be the Weyl formula~: 
$\Di \; N(\lambda )\; \sim _{+\infty} \; \frac{\lambda}{4\pi} |\mM |\; .$ \\ 
S. Gol\'enia and S. Moroianu in 
\cite{Go-Mo} have such examples.

In the case of the Poincar\'e half-plane, $\mM =\bH \; ,$ we prove 
in \cite{Mo-Tr} that 
the Weyl formula is not  valid~: 
$\Di \lim_{\lambda \to +\infty} \lambda ^{-1} N(\lambda )\; = \; +\infty \; .$\\ 
For example when $\Di \ \mb (z)=a_0^2(x/y)^{2m_0}+a_1^2y^{m_1} 
+a_2^2/y^{m_2}\; ,\ a_j>0$ and $m_j\in \bN^\star \; ,$ then  
$$ N(\lambda )\; \sim _{+\infty}\; 
\lambda^{1+1/(2m_0)} \ln (\lambda ) \alpha (m_0,m_1,m_2)\; .$$ 
\indent 
In this paper, we are interested by the hyperbolic surfaces with infinite area. 
When $\mM $    is a geometrically finite hyperbolic 
surface of infinite area and  when the above example  is arranged 
for this new situation, ($m_0$ is absent, $m_1$ appears in the cusps 
and $m_2$ in the funnels), we get  
$$ N(\lambda )\; \sim _{+\infty}\; 
\lambda^{1+1/m_2}  \alpha (m_2)\; :$$  
the  cusps do not contribute to the leading part of $N(\lambda )\; .$

\section{Main result} 

We assume that $(\mM ,\; g)\; $ is a smooth connected Riemannian manifold 
of dimension two, which is a geometrically finite hyperbolic surface 
of infinite area; 
  (see \cite{Per} or \cite{Bor} for the definition 
and the related references). More precisely 
\begin{equation}\label{hyM} 
\mM \; =\; \left (\bigcup_{j=0}^{J_1}M_j\right ) \bigcup 
\left ( \bigcup_{k=1}^{J_2} F_k\right ) \; ; 
\end{equation} 
where the $M_j$ and the $F_k$ are open sets of \mM , such 
that the closure of $M_0$ is compact, and if $J_1>0\; ,$ the other 
$M_j$ are cuspidal ends of \mM ,  and the 
$F_k $ are funnel ends of \mM . 

This means that, for any $j,\ 1\leq j\leq J_1\; ,$ there exist strictly positive constants $a_j\ \MR{and}\ L_j$ such that $M_j\; $ is isometric to 
$\bS \times ] a_j^2, +\infty [\; ,$ equipped with the metric 
\begin{equation}\label{gCups} 
 ds_j^2\; =\; y^{-2} (\ L_j^2 \ d\theta^2\; +\; dy^2\ )\; ;
\end{equation}

$(\Di  \bS \; =\; \bS^1\; $   is the unit circle.) \\  

 In the same way, for any $k,\ 1\leq k\leq J_2\; $,there exist strictly positive constants
$\Di \alpha_k\ \MR{and}\ \tau_k\ $such that $\ F_k\; $ is isometric to 
$\bS \times ] \alpha_k^2, +\infty [\; ,$ equipped with the metric 
\begin{equation}\label{gFunnel} 
ds_k^2\; =\; \tau_k^2 \cosh^2(t)  d\theta^2\; +\; dt^2 \; ;
\end{equation} 

moreover, for any two integers 
$j,\ k\ >\; 0\; ,$ we have$\   M_j\cap F_k\; =\; \emptyset \; $ and \\ 
$ M_j\cap M_k\; =\; F_j\cap F_k\; =\; \emptyset $ if $j\neq k \; .$

Let us choose some $z_0\; \in \; M_0\; $ and let us define 
\begin{equation}\label{Dd} 
d\; :\; \mM \; \to \; \bR_+\; ;\quad d(z)\; =\; d_g(z,z_0)\; ; 
\end{equation} 
$d_g (\; .\; ,\; .\; )\; $ denotes the distance with respect to the metric
 $g$.

We assume the smooth one-form 
$\; A\; $ to be given such that the magnetic field  $\mbt \; $ satisfies 
\begin{equation}\label{Hmag0} 
\lim_{d(z)\to \infty} \; \mb (z)\; =\; +\infty \; . 
\end{equation} 
If $J_1\;>\; 0\; ,$ there exists a constant 
$C_1\; >\; 0\; $ such 
\begin{equation}\label{Hmag1} 
|X\mbt (z)|\; \leq  \; C_1(\mb (z)\; +\; 1)e^{d(z)} |X|_g\; ; 
\end{equation} 
$$
\forall \; z\; \in \; M_j\; ,\ 
\forall \; X\; \in \; T_z\mM \; \ 
\MR{and}\ \forall \; j=1,\ldots J_1\; . 
$$
There exists a constant 
$C_2\; >\; 0\; $ such 
\begin{equation}\label{Hmag2} 
|X\mbt (z)|\; \leq  \; C_2(\mb (z)\; +\; 1)|X|_g \; ; 
\end{equation} 
$$ 
 \forall \; z\; \in \; F_k\; ,\ 
\forall \; X\; \in \; T_z\mM \; \ 
\MR{and}\ \forall \; k=1,\ldots J_2\; . 
$$

For any self-adjoint  operator $P\; ,$ and for any 
real $\lambda \; ,$ we will denote by $E_\lambda (P)$ its spectral projection, 
and   when  its trace is finite we will denote it by 
$$ N(\lambda ; P)\; =\; Tr(E_\lambda (P))\; .$$ 
$ N(\lambda ; P)$  is the number of eigenvalues of 
$P\; ,$ (counted with their multiplicity), which are in $]-\infty , \lambda [\; .$ 

\begin{theorem}\label{magn} 
Under the above assumptions, 
  $\ -\Delta_A \; $ has a compact resolvent and for any 
$\delta \; \in \; ]\frac{1}{3}, \frac{2}{5}[\; ,$  
 there exists a constant $\; C\; > \; 0\; $ such that 
 $$ 
 \frac{1}{2\pi} \int_{\mM} (1-\frac{C}{(\mb (m)+1)^{(2-5\delta)/2}}) 
  \ \cN (\lambda (1-C\lambda^{-3\delta +1}) -\frac{1}{4} , \mb (m)) 
\; dm \; 
 $$  
 \begin{equation}\label{magnE} 
 \leq \; N(\lambda , -\Delta_A)\; \leq   
 \end{equation} 
 $$  
  \frac{1}{2\pi} \int_{\mM} (1+\frac{C}{(\mb (m)+1)^{(2-5\delta)/2}}) 
 \  \cN (\lambda (1+C\lambda^{-3\delta +1}) -\frac{1}{4} , \mb (m))  
\; dm \;  
$$
where 
$$ \cN ( \mu , \mb (m)) \; =\;  \mb (m) 
\sum_{k=0}^{+\infty} [\mu 
 - (2k+1) \mb (m)]_{+}^{0} \;\quad \mbox{if}\ 
\  \mb (m)\; >\; 0\; ,$$ and 
$$\cN ( \mu , \mb (m)) \; =\;  \mu/2\; \quad \mbox{if}  \ \ \Di \mb (m)\; =\; 0\; .$$ \\ 
$[ \rho ]^{0}_{+}$ is the Heaviside function: 
$$[ \rho ]^{0}_{+}\; =\; \left\{ \begin{array}{ccc} 1\; ,& \MR{if} & \rho > 0\\ 
0\; , & \MR{if} & \rho \leq 0\; .
\end{array} 
\right . $$ 
\end{theorem} 

The Theorem remains true if we replace
$\Di \int_{\mM} \  $ by $\Di \ \ \sum_{k=1}^{J_2} \int_{F_k} \; $, due to the fact that the other parts are bounded by $C\lambda \; .$

\begin{corollary}\label{magnC} 
Under the assumptions of Theorem \ref{magn} 
and if the function 
$$\omega (\mu )\; =\; \int_{\mM}  [\mu - \mb (m)]_+^0 dm
$$ 
satisfies, $\Di \  \exists \; C_1 > 0\ \MR{s.t.}\ \forall \; \mu > C_1\; ,\ 
\forall \; \tau \; \in \; ]0,1[\; ,$ 
\begin{equation}\label{hypW} 
\omega\ ( (1+\tau )\ \mu) - \omega (\mu ) \leq   C_1 \ \tau\ \omega (\mu ) \; , 
\end{equation} 
then  
\begin{equation}\label{magne} 
N(\lambda ; -\Delta_{A})\; \sim \;  \frac{1}{2\pi}
\int_{\mM } 
\cN (\lambda -\frac{1}{4}, \mb (m))\; dm\; . 
\end{equation} 

\end{corollary} 

For example this allows us to consider  magnetic fields  of the following type:  
\begin{center} 
on $\Di \ F_k\; ,\ 
\; \mb (\theta ,t)\; =\; p_k(\ 1/\cosh (t)\ )\; ,$ \\ 
and on $\Di \ M_j\; ,\ j>0\; , \ \mb (\theta ,y)\; =\; q_j(y)\; , $ 
\end{center}  
where the 
  $p_k(s)$ and the $q_j(s)$ are,
for large $s$,   polynomial functions of order $\geq 1\ .$ 
In this case, if $d$ is the largest order of the $p_k(s)\; ,$ then 
$$N(\lambda ; -\Delta_{A})\; \sim \; \alpha  \lambda ^{1+1/d}\; ,$$ 
for some constant $\alpha > 0\; ,$ depending only on the funnels 
$F_k$ where the order of $p_k(s)$ is $d\; .$

\section{Estimate for  Dirichlet operators} 

\subsection{The main propositions} 

In this section, we consider some particular open set $\Omega $ 
of $\mM $ with smooth boundary. To $\Omega $ and 
$-\Delta_A\; ,$   we associate 
the Dirichlet operator $-\Delta_A^\Omega\; ,$ 
and we estimate $N(\lambda ; -\Delta_A^\Omega )\; .$ 

\begin{proposition}\label{Dir0} 
Let $\Omega $ an open set of $M_0$ with smooth boundary.  
Then there exists a constant $C_\Omega >0$ s.t. 
$$\left | N(\lambda ; -\Delta_A^\Omega ) \; 
-\; \frac{|\Omega |}{4\pi }\lambda \right |\; \leq \; C_\Omega \sqrt{\lambda} 
\; ;\quad \forall \; \lambda > 1\; .$$ 
\end{proposition} 

As $\Di \overline{\Omega }$ is compact, the above estimate is well known.
See for example \\ Theorem 29.3.3 in \cite{Hor}. 

\begin{proposition}\label{DirCusp} 
Let $j>0$ and $\Omega $ an open set of the cusp $M_j$, isometric 
to  
$\bS \times ] a^2, +\infty [\; ,$ equipped with the metric 
$$ 
 ds^2\; =\; y^{-2} (\ L^2 \ d\theta^2\; +\; dy^2\ )\; ;\quad 
(a\ \MR{and}\ L\ \MR{are\ strictly\ positive \ constants})\; .
$$  
Then $\ -\Delta_A^\Omega \; $ has a compact resolvent and 
$$ N(\lambda ; -\Delta_A^\Omega ) \; 
\sim \; \frac{|\Omega |}{4\pi }\lambda  \; ; \quad \MR{as}
\quad \lambda \to +\infty 
\; .$$ 
\end{proposition} 
 We will prove it in the next subsection. 

\begin{proposition}\label{DirFun} 
Let  $\Omega $ an open set of a funnel $F_k$, isometric 
to  
$\bS \times ] a^2, +\infty [\; ,$ equipped with the metric 
$$ 
 ds^2\; =\; L^2 \cosh^2(t)\ d\theta^2\; +\; dt^2\; ;\quad 
(a\ \MR{and}\ L\ \MR{are\ strictly\ positive \ constants})\; .
$$  
Then 
 $\ -\Delta_A^\Omega \; $ has a compact resolvent and for any 
$\delta \; \in \; ]\frac{1}{3}, \frac{2}{5}[\; ,$  
 there exists a constant $\; C\; > \; 0\; $ such that 
 $$ 
 \frac{1}{2\pi} \int_{\Omega } (1-\frac{C}{(\mb (m)+1)^{(2-5\delta)/2}}) 
  \ \cN (\lambda (1-C\lambda^{-3\delta +1}) -\frac{1}{4} , \mb (m)) 
\; dm \; 
 $$  
 $$
 \leq \; N(\lambda , -\Delta_A^\Omega )\; \leq   
 $$
 $$  
  \frac{1}{2\pi} \int_{\Omega} (1+\frac{C}{(\mb (m)+1)^{(2-5\delta)/2}}) 
 \  \cN (\lambda (1+C\lambda^{-3\delta +1}) -\frac{1}{4} , \mb (m))  
\; dm \;  
$$

\end{proposition} 
The proof comes easily following the ones in the Poincar\'e 
half-plane of \cite{Mo-Tr}, using the method of \cite{Col}, in the neighbourhood of the boundary 
at infinity. It corresponds to a context where the partitions of unity were fine, so they can be performed on $\bS \times ] a^2, +\infty [\; ,$ 
(instead of $\bR \times ]  - \infty , 0 [\; )\; .$

\subsection{Proof of Proposition \ref{DirCusp} } 

For simplicity we change the unit circle 
$\bS =\bS_1$ into the circle $\bS_L\; ,$ of radius $L\; ,$  
so  
\begin{equation}\label{SLeq} 
\Omega =\bS_L\times ]a^2 , +\infty [\; ,\quad ds^2\; =\; y^{-2} ( dx^2 \; +\; dy^2 )\; ,\quad 
\MR{and} 
\end{equation} 
$$\quad 
-\Delta_A^\Omega u(z)= y^2[(D_x -A_1)^2u(z) +(D_y -A_2)^2 u(z)]\; ; $$
moreover $\Di \ d(z,z')=\arg \cosh \frac{y^2 +y'^2+d^{2}_{\bS_L} (x,x')}{2yy'}\; .$

We begin by proving the compactness of the resolvent of 
$-\Delta_A^\Omega\; .$ 
\begin{lemma}\label{bCuspL} 
There exists $C_0 > 1$ such that 
$$\int_\Omega (\mb (z) -C_0) |u(z)|^2 dm \; \leq \; \int_\Omega 
-\Delta_A^\Omega u(z)\overline{u(z)}dm\; ;\quad \forall \; 
u\; \in \; C^\infty_0(\Omega )\; . $$
\end{lemma} 
{\bf Proof.} Let us denote the quadratic form 
\begin{equation}\label{defq} 
q_A^\Omega (u)\; =\; \int_\Omega 
-\Delta_A^\Omega u(z)\overline{u(z)}dm\; \quad \forall \; 
u\; \in \; C^\infty_0(\Omega )\; . 
\end{equation} 
Then $\Di q_A^\Omega (u)\; =\; \int_\Omega 
\left [ |(D_x -A_1)u|^2\; +\; (D_y -A_2)u|^2 \right ] dxdy\; ,$\\ 
and  $\Di \left | \int_\Omega \mbt (z) |u(z)|^2 dm\; \right | $\\
$$
=\; \left | \int_\Omega [ (D_x -A_1)u(z)\overline{(D_y -A_2)u(z)} 
\; -\; (D_y -A_2)u(z)\overline{(D_x -A_1)u(z)} ] dx dy \; \right |
\; .$$ 
Therefore we get that $\Di \left | \int_\Omega \mbt (z) |u(z)|^2 dm\; \right |\; \leq \; q_A^\Omega (u)\; .$\\ 
As $\mb (z)=|\mbt (z)|\; \to \; +\infty $ at the infinity, 
the Lemma comes easily. 

The Lemma \ref{bCuspL} and the assumption (\ref{Hmag0}) prove 
that $-\Delta_A^\Omega $ has compact resolvent. 

Later on, we will need that the assumptions (\ref{Hmag0}) 
and   (\ref{Hmag1}) ensure 
that there exists $C>1$ such that 
$\Di \ \forall \; z=(x,y)\; ,\ z'=(x',y')\ \in \; \Omega \; ,$ 
\begin{equation}\label{compB} 
\mb (z)/C\; \leq \; \mb (z')\; \leq \; C\mb (z)\; , \quad 
 \MR{if}\ 
|y-y'|\leq 1\ \MR{and}\  y> C\; . 
\end{equation} 
This comes from the fact that $d(z)$ is equivalent to $\ln (y)$ for $y(>1)$ large enough, so 
the assumption (\ref{Hmag1}) ensures that $|\partial_x b(z)|+|\partial_y b(z)|\leq C (|b(z)|+1). $

\begin{lemma}\label{cuspR} There exists a constant $C_0>1$ such that, for any $\lambda > 1$ and 
for any  $K\subset \Omega $ isometric to 
$I_1\times I_2\; ,$ endowed with the metric in (\ref{SLeq}), 
with 
$$I_1=]x_0 -\epsilon_1, x_0+\epsilon_1[\; ,\ 
I_2=]y_0 -\epsilon_2, y_0+\epsilon_2[\; ,$$ 
$$ \epsilon_1 \; \in \; ]C_{0}^{-1},1[\; ,
\ \epsilon_2= \sqrt{y_0}/\sqrt{\mb (z_0)}\; , \ (y_0>C_0)\; ;$$ 
the following estimates hold: 
\begin{equation}\label{NLK} 
[\lambda (1-\frac{1}{\sqrt{y_0}})  -C_0]\frac{|K|_g}{4\pi} 
\; \leq \; N(\lambda ; -\Delta_A^K )\; 
\leq \; 
  [\lambda (1+\frac{1}{\sqrt{y_0}})  +C_0]\frac{|K|_g}{4\pi}\; .
\end{equation} 
\end{lemma}

{\bf Proof.} If $\mb (z_0) > C\lambda \; ,$ then, 
according to the estimate of Lemma \ref{bCuspL} 
with $K$ instead of $\Omega \; ,\quad N(\lambda ; -\Delta_A^K)\; =\; 0\; .$\\ 
So we can assume that $\ \mb(z_0)\leq C\lambda \; .$

We use that the spectrum of $-\Delta_A^K$ 
is gauge-invariant, so we can suppose that in 
$K$ 
$$A_2=0\quad \MR{and}\quad A_1(x,y)=-\int_{y_0}^{y} \frac{\mbt (x,\rho)}{\rho^2} d\rho \; .$$ 
Then $\Di \ |A_1(x,y)|\leq C\epsilon_2\frac{\mb (z_0)}{y_0^2} \; .$\\ 
From this estimate, we get that for any 
$\epsilon \in ]0,1[\; ,$ 
$$-(1-\epsilon)\Delta_0^K \; -\; C\epsilon_2^2
\frac{\mb^2 (z_0)}{\epsilon y_0^2}\; \leq \; 
-\Delta_A^K \; \leq \; 
-(1+\epsilon)\Delta_0^K \; +\; C\epsilon_2^2
\frac{\mb^2 (z_0)}{\epsilon y_0^2}\; .$$ 
We take $\; \epsilon =1/\sqrt{y_0}\; ,$ to get 
$$-(1-\frac{1}{\sqrt{y_0}}  )\Delta_0^K \; -\; C
\frac{\mb  (z_0)}{\sqrt{ y_0} }\; \leq \; 
-\Delta_A^K \; \leq \; 
-(1+\frac{1}{\sqrt{y_0}})\Delta_0^K \; +\; C
\frac{\mb (z_0)}{\sqrt{ y_0} }\; .$$
As $\mb(z_0)\leq C\lambda \; ,$ the Lemma follows easily 
from the min-max principle and the well-known estimate for  
$N(\lambda ; -\Delta_0^K )\; .$

{\bf Proof of Proposition \ref{DirCusp}.} 

It follows easily from Lemma \ref{cuspR}, (for large $ y\; ),$ 
using the same tricks as in \cite{Mo-Tr}.

\section{Proof of the main Theorem \ref{magn}} 

The proof comes easily from the three  propositions 
\ref{Dir0} - - \ref{DirFun}, following the method developped
 in \cite{Mo-Tr}.

\section{Remark on the case of constant magnetic field} 

It is not always possible to have a constant magnetic field 
on $\mM\; ,$ (for topological reason), but for 
any $\Di (b ,\beta )\; \in \; \bR^{J_1}\times \bR^{J_2}\; ,$ 
there exists a one-form $\ A\; ,$ such that the corresponding magnetic field 
$dA\; $ satisfies
\begin{equation}\label{asymptC} 
dA\; =\; \mbt (z) dm\;\quad 
\left \{ \begin{array}{c} 
\mbt (z)\; =\; b_j\ \forall\; z\; \in \;  M_j\\ 
\mbt (z)\; =\; \beta_k\ \forall\; z\; \in \;  F_k 
\end{array} 
\right . 
\end{equation}

\begin{theorem}\label{ThC} 
 Assume (\ref{hyM}) and (\ref{asymptC}). 

If $J_1\; =\; 0\; $ and $J_2\; >\; 0\; ,$  then the essential spectrum 
of $\Di \; -\Delta_A\; $ is 
\begin{equation}\label{FThC} 
\sp_{ess} (-\Delta_A)\; =\; 
[\frac{1}{4}+\inf_k \beta_k^2 \; ,\  +\infty [ 
\, \bigcup \left ( \bigcup_{k=1}^{J2} S(\beta_k) \right )  
\end{equation}
with $S(\beta_k)=\emptyset $ when $|\beta_k |\leq 1/2\; $ 
and when $|\beta_k | > 1/2\; $\\ 
$\Di S(\beta_k)=\{ (2j+1)|\beta_k| -j(j+1)\; ; \ j\; \in \; \bN ,
\ j < |\beta_k| -1/2 \} \; .$  

If $J_1$ and $J_2$ are $ >\; 0\; ,$ then for any $j\; ,\ 1\leq j\leq J_1\; $ 
and for any $\; z\; \in \; M_j\; $ 
there exists a unique closed curve through $\Di \; z\; ,\   
 \cC_{j,z}\; $ in $\ (M_j, \ g)\; ,$ 
not contractible and  with zero $g-$curvature. 
The following limit 
exists and is finite: 
\begin{equation}\label{Aj} 
[A]_{M_j}\; =\; \lim_{d(z)\to +\infty} \; \int_{\cC_{j,z}} 
A\; . 
\end{equation} 

If $J_1^A\; =\; \{ j\in \bN \; , \ 1\leq j\leq J_1\ s.t.\ 
[A]_{M_j}\in 2\pi \bZ\; \} \; ,$ then 
\begin{equation}\label{CThC} 
\sp_{ess} (-\Delta_A)\; =\; 
[\frac{1}{4}+\min \{ \inf_{j\in J_1^A} b_j^2\; ,\; 
\inf_{1\leq k \leq J_2} \beta_k^2\; \} \;  , \ +\infty [
\, \bigcup \left ( \bigcup_{k=1}^{J2} S(\beta_k) \right ) \; . 
\end{equation} 

If $J_2=0$ and $J_1^A\; =\; \emptyset \; ,$ then 
$\ \sp_{ess} (-\Delta_A)\; =\; \emptyset \; :$\\ 
$-\Delta_A\; $ has purely discrete spectrum, (its resolvent 
is compact).

\end{theorem}

 \begin{remark}\label{RmC} In Theorem \ref{ThC},
 one can change $\cC_{j,z}$ into $\cS_{j,z}\; ,$
 the unique closed curve through $z\; ,$ not contractible and with
 minimal $g-$length.\\
 $\cS_{j,z}\; $ is not smooth at $z,\ \cS_{j,z}\; $ is part
 of two geodesics through  $z\; ,$ so there is an out-going tangent
 and an incoming tangent at $z\; .$
  It is easy to see that
 $\Di \; \cC_{j,z} \; \cap\;  \cS_{j,z}\; =\; \{ z\} \; ,$
 so by Stokes formula
 $$\Di \; \int_{\cS_{j,z}} (A -A^0)\;
 =\; \int_{\cC_{j,z}} (A-A^0)\; ,$$
 where $A^0\; $ is a one-form on $M\; ,$ such that
 $$dA\; =\; dA^0\quad
 \MR{on}\ \ M_j \ \  \MR{and}\quad [A^0]_{M_j}\; =\; 0\; ;\
 \forall \; j\; .$$
 \indent
 The orientation in both cases $ \cC_{j,z}\; $ and $\cS_{j,z}\; ,$
 is chosen such that, if \\
 $u_z,\ v_z\ \in T_zM_j\; ,\ g_z(u_z,v_z)=0\; ,\ dm(u_z,v_z)>0\; ,\
 $and  $u_z$ is tangent to the curve (in the positive direction), then
 $v_z$ points to boundary at infinity;
 (for $\cS_{j,z}\; ,$ one can take as $ u_z\; $ the
 out-going tangent, or
 the incoming tangent).
 \end{remark}
 
{\bf Proof of Theorem \ref{ThC}.} 
It is clear that 
\begin{equation}\label{spEss} 
\sp_{ess} (-\Delta_A)\; =\; \left ( \bigcup_{j=1}^{J_1} 
\sp_{ess} (-\Delta_{A}^{M_j})\right ) \bigcup 
\left ( \bigcup_{k=1}^{J_2} 
\sp_{ess} (-\Delta_{A}^{F_k})\right )\; ; 
\end{equation} 
so the proof will result  on the two lemmas below.
\begin{lemma}\label{Fspess} 
$$ \sp_{ess} (-\Delta_{A}^{F_k})\; =\; 
[\frac{1}{4} + \beta_k^2 \; , \ +\infty [
\, \cup\, S(\beta_k)\;  \; . $$ 
\end{lemma} 
{\bf Proof.} 
  We have $\Di -\Delta_{A}^{F_k}\; =\; 
\tau_{k}^{-2} \cosh^{-2}(t)(D_\theta -A_1)^2 
+ \cosh^{-1}(t) (D_t -A_2) \left[\cosh(t) (D_t -A_2) \right]\; .$\\ 
Since $\Di \mbt =\beta_k =\tau_{k}^{-1}\cosh^{-1}(t)
(\partial_\theta A_2 -\partial_t A_1)\; ,$ there exists a function $ \varphi$
such that  
$\Di A - {\widetilde A}\; =\; d\varphi\ \MR{if}\ 
{\widetilde A}=(\xi -\beta_k \tau_k\sinh (t))d\theta \; ,\ (\MR{for\ some\ constant}\ \xi )\; .$\\   
So we can assume that $A={\widetilde A}\; .$ 

We change the density $dm=\tau_k \cosh (t) d\theta dt\; $ for  
$d\theta dt\; ,$ using the unitary operator 
$\Di Uf=(\tau_k \cosh (t))^{1/2}f\; ,$ so 
$$P=-U\Delta_{A}^{F_k}U^\star = 
\tau_{k}^{-2} \cosh^{-2}(t)(D_\theta -A_1)^2 
+  D_t^2 +\frac{1}{4}(1+\cosh^{-2}(t))\; .$$ 
We remind that  $\lambda \; \in \; \sp_{ess} (-\Delta_{A}^{F_k})\; $ 
iff there exists a sequence 
$(u_j)_j\; \in \; Dom(-\Delta_{A}^{F_k})$ converging weekly 
in $L^2(F_k)$ to zero,  
$\displaystyle \| u_j\|_{L^2(F_k)} =1$ 
and such that the sequence $(-\Delta_{A}^{F_k}u_k -\lambda u_k)_k$ 
converges strongly to zero. 

It is clear that   
$\Di \sp (-\Delta_{A}^{F_k})\; =\; \sp (\bigoplus_{\ell \in \bZ} 
\; P_\ell )\; ,$\\ 
$$\ P_\ell 
=D_t^2 + \tau_{k}^{-2} \cosh^{-2}(t)(\ell +\beta_k \tau_k 
\sinh (t) -\xi )^2  +\frac{1}{4}(1+\cosh^{-2}(t))\; ,$$ 
for the Dirichlet condition on $L^2(I; dt)\; ;\ 
I=]\alpha_k^2\; ,\  +\infty [\; .$\\ 
So  $\displaystyle \ \sp (-\Delta_{A}^{F_k})=\bigcup_{\ell \in \bZ} 
\sp (P_\ell )\; .$

Writing that $\Di P_\ell \; =\; D_t^2 + 
\left ( \frac{\ell -\xi}{\tau_k \cosh (t)} +\beta_k\tanh (t) 
\right )^2 +\frac{1}{4}(1+\cosh^{-2}(t)) \; ,$\\ 
we get easily that $\Di \sp_{ess} (P_\ell )=[ \frac{1}{4} +\beta_k^2, +\infty[ 
\; ,$  
and that the number of eigenvalues $\Di < \frac{1}{4} + \beta_k^2$ is finite for 
all $\ell <\xi $ and equal to zero for all $\ell \geq \xi \; . $ 
Here we assume $\beta_k >0\; .$   
So $\displaystyle \ [  \frac{1}{4} +\beta_k^2, +\infty[\; \subset \; 
\sp_{ess} (-\Delta_{A}^{F_k})\; $  
and the other part of $\sp_{ess} (-\Delta_{A}^{F_k})$ is 
$\displaystyle S_\infty =\{ \lambda \; ;\ \lambda =\lim_{j\to +\infty } 
\lambda_{\ell (j)} \; ,\ \lambda_{\ell (j)}\in \sp_d (P_{\ell (j)}  )\} \ $,\\ 
where $ (\ell (j))_j$ denotes any decreasing sequence of negative integers.\\ 
Now we use again the formula\\
 $\displaystyle P_\ell \; =\; D_t^2 + 
\left ( \frac{\ell -\xi}{\tau_k \cosh (t)} +\beta_k\tanh (t) 
\right )^2 +\frac{1}{4}(1+\cosh^{-2}(t)) \; .$\\
 Assuming $\ell -\xi < 0\; ,$ we set 
$ \rho = |\ell -\xi |/\tau_k$ and we introduce the new variable
$y=2\rho e^{-t}\; .$ We get that $P_\ell $ is unitarily equivalent 
to $\Di {\widetilde P}_\rho $ defined as a Dirichlet type  
operator 
in $L^2(]0,2\rho e^{-\alpha_k^2}[; dy)\; ,$ (zero boundary 
condition is only required on the right boundary): 
$${\widetilde P}_\rho \; =\; D_y (y^2 D_y)  
+ W_\rho (y)\; ,\quad {\rm with}$$
$$\quad W_\rho (y)= \left ( 
\beta_k\frac{(1-y^2/(4\rho^2)}{1+y^2/(4\rho^2)} -\frac{y}{1+y^2/(4\rho^2)} \right )^2 \; 
+\; \left(\frac{y/(2\rho)}{1+y^2/(4\rho^2)}\;\right)^2 \; .$$ 
So we have  $\Di \lim_{\rho \to +\infty } W_\rho (y)=W_\infty (y)=(\beta_k -y)^2\; ,$
and the operator \\$\Di {\widetilde P}_\infty = D_y (y^2 D_y) 
+ W_\infty (y)\; $ on 
$L^2(]0,+\infty [; dy )$ 
satisfies, (see \cite{Mo-Tr} ), \\ $\sp ({\widetilde P}_\infty ) 
= \sp_{ess}  ({\widetilde P}_\infty ) \; \cup \; 
\sp_d ({\widetilde P}_\infty )$ with 
$$\sp_{ess}  ({\widetilde P}_\infty ) \; =\; [  \frac{1}{4} +\beta_k^2 , +\infty [ 
\; ;\quad \quad \sp_d ({\widetilde P}_\infty )\; =\; S(\beta_k)\; . $$ 
We remind that the eigenfunctions associated to the eigenvalues 
in $S(\beta_k)$ of $\Di {\widetilde P}_\infty $ are 
exponentially decreasing, so if 
 $\lambda_0 (\rho )\leq \ldots \leq  \lambda _j(\rho )\leq \lambda _{j+1} (\rho )\ldots $ \\ 
are the eigenvalues of $\Di {\widetilde P}_\rho $ then 
for any $\Di j\; ,$\\ 
$\Di  \lim_{\rho \to +\infty } \lambda _j(\rho ) 
=\lambda _j(\infty  )=(2j+1)\beta_k -j(j+1)\; ,$ 
if $\beta_k > 1/2$ and $j < \beta_k -1/2\; ,$ 
otherwise $\Di \lim_{\rho \to +\infty } \lambda _j(\rho )= \frac{1}{4} + \beta_k^2\; .$ \\ 
Therefore we get that $S_\infty \; =\; S(\beta_k)\; , $ 
or  $S_\infty \; =\; S(\beta_k)\; \cup \; \{   \frac{1}{4} +\beta_k^2 \} \; :$ 
the formula of Lemma \ref{Fspess} follows.

\begin{lemma}\label{Cspess} If $1\leq j\leq J_1\; $ and 
$j\; \notin \; J_1^A\; ,$ then 
$$ \sp_{ess} (-\Delta_{A}^{M_j})\; =\; \emptyset 
\; . $$ 
\indent 
If $j\; \in \; J_1^A\; ,$ then 
$$ \sp_{ess} (-\Delta_{A}^{M_j})\; =\; 
[\frac{1}{4} + b_j^2 \; ,\  +\infty [\; . $$ 

\end{lemma} 
{\bf Proof.}
 Use the coordinate $t=\ln y\; $ instead of $y\; ,$ so 
$$M_j\; =\; \bS \times ]\alpha_j^2 , +\infty [\;  \quad \MR{and}\quad ds_j^2\; =\; L_j^2e^{-2t} d\theta^2 \; +\; dt^2\; ;\ \ (\alpha_j=e^{a_j} )\; .$$
Then  $\Di -\Delta_{A}^{M_j}\; =\; 
L_{j}^{-2} e^{2t}(D_\theta -A_1)^2 
+ e^t (D_t -A_2)(e^{-t}(D_t -A_2))\; ,$\\ 
$\Di \mbt =\ L_{j}^{-1}e^t
(\partial_\theta A_2 -\partial_t A_1)\; $ and 
$\ dm=L_j e^{-t} d\theta dt\; .$ As in Lemma \ref{Fspess}, 
we have  
$$A - {\widetilde A}\; =\; d\varphi\ \MR{if}\ 
{\widetilde A}=(\xi +L_jb_je^{-t})d\theta \; ,\ (\MR{for\ some\ constant}\ \xi )\; .$$  
So we can also assume that $A={\widetilde A}\; .$ 

We replace the density $dm\; $ by
$d\theta dt\; ,$ using the unitary operator\\  
$\Di Uf=\sqrt{L_j}  e^{-t/2}f\; ,$ so 
$$P=-U\Delta_{A}^{M_j}U^\star = 
L_{j}^{-2} e^{2t}(D_\theta -A_1)^2 
+  D_t^2 +\frac{1}{4}\; .$$ 
Then we get also that  
$$\sp (-\Delta_{A}^{M_j})=\sp (P)=\bigcup_{\ell \in \bZ} 
\sp (P_\ell )\; ;\ P_\ell 
= D_t^2 + \frac{1}{4} + \left ( e^t \frac{(\ell + \xi )}{L_j} +b_j 
\right )^2 \; ,$$ 
for the Dirichlet condition on $L^2(I; dt)\; ;\ I=]\alpha_j^2, +\infty [\; .$ 

When $\ell +\xi \neq 0\; ,$ the spectrum of $P_\ell $ 
is discrete.  More precisely 
$$\sp (P_\ell )\; =\; \sp (P^{\pm})\; ,\quad 
\mbox{where} \quad P^{\pm } =D_t^2+\frac{1}{4}+(\pm e^t+b_j)^2$$ 
for the Dirichlet condition on $L^2(I_{j,\ell}; dt)\; $;$\ I_{j,\ell}=]\alpha_j^2 +\ln (|\ell +\xi |/L_j), +\infty [\; ,$ 
and $\Di \pm =\frac{\ell +\xi }{|\ell +\xi |} \; .$  

So $\Di \lim_{|\ell |\to \infty} \; \inf 
\sp (P_\ell )\; =\; +\infty\; ,$ 
and then we get easily that the spectrum of 
$ -\Delta_{A}^{M_j}\; $ is discrete,
 when 
$\Di \xi =[A]_{M_j}/(2\pi )\; \notin \;  \bZ \; .$ 

If $\ell +\xi = 0\; ,$ the spectrum of $P_\ell $ is absolutely continuous~: 
$$\sp (P_{-\xi })\; =\; \sp_{ess} (P_{-\xi }) \; 
=\; \sp_{ac} (P_{-\xi })\; 
\; = \; [\frac{1}{4} 
+b_j^2 \; ,\  +\infty [\; ;$$ 
and then, when $\Di [A]_{M_j}\; \in \; 2\pi \bZ \; ,\quad 
\sp_{ess}( -\Delta_{A}^{M_j})\; =\; 
[\frac{1}{4} 
+b_j^2 \; ,\  +\infty [\; .$ \\ 
This achieves the proof of Lemma \ref{Cspess}.

\end{document}